\documentclass[aps,prl,twocolumn,showpacs]{revtex4}

\usepackage{amsmath,amssymb,latexsym}
\usepackage{graphicx}
\usepackage{amsfonts}
\usepackage{amsthm}
\usepackage{stmaryrd}
\usepackage{color}

\newcommand{\traj}{{\boldsymbol \omega}}
\newcommand{\id}{\textrm{d}}
\newcommand{\mean}[1]{{\left< #1 \right>}}

\newcommand{\prob}{{\cal P}}

\begin{document}

\title{Fluctuations and response of nonequilibrium states}
\author{Marco Baiesi}
\affiliation{Instituut voor Theoretische Fysica, K.U.Leuven, 3001,
Belgium.}
\author{Christian Maes}
\affiliation{Instituut voor Theoretische Fysica, K.U.Leuven, 3001,
Belgium.}
\author{Bram Wynants}
\email[]{Bram.Wynants@fys.kuleuven.be}
\affiliation{Instituut voor Theoretische Fysica, K.U.Leuven, 3001,
Belgium.}

\date{\today}

\begin{abstract}
A generalized fluctuation-response relation is found for thermal systems
driven out of equilibrium. Its derivation is independent of many
details of the dynamics, which is only required to be first-order.
The result gives a correction to the equilibrium
fluctuation-dissipation theorem, in terms of the correlation
between observable and excess in dynamical activity caused by
the perturbation. Previous approaches to this problem are
recovered and extended in a unifying scheme.

\end{abstract}

\maketitle

The fluctuation-dissipation theorem is a standard chapter in statistical 
mechanics~\cite{kubo,mprv,sethna}. A system in
thermal equilibrium has statistical fluctuations proportional to
its response to external perturbations: a small
impulse changing the potential $U \rightarrow U - h_s \,V$ at time
$s$, will produce a response
$R_{QV}^{\textrm{eq}}(t,s)=\delta\mean{Q(t)}/\delta h_s$
in a quantity $Q$ at time $t\ge s$ of the form
\begin{equation}\label{efd}
R_{QV}^{\textrm{eq}}(t,s) = \beta\,\frac{\id}{\id s}\mean{V(s)\,Q(t)}_{\textrm{eq}}
\end{equation}
where $\mean{V(s)\,Q(t)}_{\textrm{eq}}$ is a correlation function quantifying
the equilibrium fluctuations in absence of any perturbation,
and the proportionality constant $\beta=1/k_BT$ is the inverse temperature.
An early example of this theorem is present in Einstein's
treatment of Brownian motion, where the diffusion constant,
expressed as a velocity autocorrelation function, is found
proportional to the mobility. Other famous examples include the
Johnson-Nyquist formula for electronic white noise and the Onsager
reciprocity for linear response coefficients.

So far, approaches deriving
a fluctuation-dissipation relation (FDR) for 
nonequilibrium~\cite{vulp,deker,hanggi,ckp,lipp,diez,jona,gaw,blickle,ss} have
not found a physical unification and do not appear as
textbook material.  One reason may be that
previous work has not been seen to identify a sufficiently general
structure with a clear corresponding statistical thermodynamic
interpretation.  Today, such an interpretation has become available 
from advances in dynamical fluctuation
theory for nonequilibrium systems.

Aiming to provide a simple and general approach to FDRs, in this
Letter we put forward a FDR for nonequilibrium regimes in a
framework that may represent a unifying scheme for previous
formulations. Our main result can be found in a general formula, Eq.~\eqref{mainid}
below, which can also sometimes be rewritten as (\ref{nonefd})
or (\ref{marmainid}).

In order to go beyond equilibrium and beyond formal perturbation theory, 
it is important
to recognize in the right-hand side of \eqref{efd} the role 
of entropy production, as usual governing close-to-equilibrium
considerations. Eq.~\eqref{efd} expresses the
correlation between the dissipation represented by entropy production
(energy change divided by temperature)
and the observable $Q(t)$.
Here, for perturbations of a nonequilibrium system 
(that has already a non-vanishing set of flows and
hence a non-zero entropy production) we will rather speak of the excess of
entropy produced by the perturbation $h_s\,V$.  

The second ingredient that is essential in nonequilibrium
is still a less known quantity, called {\em traffic}~\cite{mnw} or dynamical
{\em  activity}~\cite{DA1}, first introduced in  \cite{t-symm}.
Perhaps activity has been somewhat overlooked in the past because it plays a truly
significant role only beyond linear order around equilibrium~\cite{mnw,dm}, and 
it is not part of the picture in standard irreversible thermodynamics.
Our present results will in fact further
clarify the relevance of dynamical activity.
To get a first idea of its meaning, we note that
activity often measures the frequency of transitions in a trajectory: this is
a time-symmetric aspect of dynamical fluctuations \cite{t-symm}
because it does not depend on the direction chosen to span the trajectory.
On the other hand, it is well known that
entropy production is time-antisymmetric~\cite{entr}, 
changing sign upon reversal of time
(fluxes are inverted if time is run backward).
Thus, both quantities arise naturally  as two complementary
pieces of the space-time action, as reminded below.

We consider a system in an environment at inverse temperature $\beta$ 
(with $k_B=1$),
where a time-independent nonequilibrium condition can be imposed by external driving
fields or by installing mechanical displacements or chemical gradients
at the boundaries of the system.
Moreover, the system evolves according to a Markovian/first-order dynamics: for any two
observables $f$ and $g$,  their correlations at times $s<t$ satisfy
$\frac{\id}{\id t}\left< f(x_t)\,g(x_s) \right> = \left<
(Lf)(x_t)\,g(x_s) \right>$ where $L$ is a linear operator, called
generator, acting on observables, 
and $x_t$ is the state at time $t$ at some reduced, e.g. mesoscopic level
of description: $e^{t L}f(x) = \mean{f(x_t)|x_0=x}$, $t\ge 0$
(in the sequel we abbreviate $f(x_t)=f(t)$).
The generator $L$ is a standard tool: for jump processes
it is the matrix with transition rates as off-diagonal elements, 
while
$L = \text{force}\cdot \nabla  + \Delta / \beta$ for overdamped diffusions.
However, it is our aim to leave the generator $L$ as unspecified as possible,
to emphasize the generality of the results, referring e.g.~to \cite{hanggi} 
for further background.

Our formulation is based on the point of view that the system can be
described by a distribution ${\cal P}(\traj)$ over its possible
trajectories $\traj$ (space-time paths), giving the state $\omega_s$
of the system at each time $s\le t$.  In
general $\prob(\traj) =  \prob(\traj|\omega_0)\,\mu(\omega_0)$, namely
$\prob(\traj)$ is the probability of the initial state  $\omega_0$,
i.e.~$\mu(\omega_0)$, times the probability of the trajectory
$\prob(\traj|\omega_0)$ given the starting point.  The ensemble
averages of a quantity $O(\traj)$ are then $\mean{O(\traj)}  \equiv
\sum_\traj \prob(\traj) O(\traj)$ where $\sum_\traj$ is a simple
notation including also a continuum of trajectories. Here we are 
interested in comparing probabilities of
trajectories for $h_s\ne 0$ (for $s\ge 0$), $\prob^h(\traj) =
\prob^h(\traj|\omega_0)\mu(\omega_0)$, with those of the unperturbed
system.  In analogy with the $h=0$ case, we use the notation
$\mean{O(\traj)}^h  \equiv \sum_\traj \prob^h(\traj) O(\traj)$ for
averages in the perturbed system.  We start by writing 
 \begin{equation}\label{Ph}
 \prob^h(\traj)  = e^{-A(\traj)}  \prob(\traj)
\end{equation}
 and by focusing on the action $A(\omega)  =
-\ln[\prob^h(\traj)/\prob(\traj)]$.
In this set-up we put ourselves in line with the
Onsager-Machlup approach but outside equilibrium, 
possibly beyond quadratic or diffusive approximations and also dealing with
jump processes. 

We ask what determines the path-space distribution
and the space-time local action governing it.
To find an answer, 
it is useful to decompose $A(\traj) = (T(\traj) - S(\traj))/2$
in terms of its time-antisymmetric component
$S(\traj) \equiv A(\theta\traj)-A(\traj)$
and the time-symmetric $T(\traj) \equiv A(\theta\traj)+A(\traj)$,
where a time-reversed state
$(\theta \omega)_s$ is equal to $\omega_{t-s}$ (with reversed momenta when applicable).
 In the time interval $[0,t]$, in standard physical situations one finds that
 \begin{equation}\label{ef}
S(\traj) =  \beta\!\left[
h_t V(t)-h_0 V(0) - \!\int_0^t \!\id s \frac{\id h_s}{\id s}V(s)\right]
\end{equation}
is the excess entropy flux from the system to its
environment. Excess is always meant in the sense of the
perturbed process with respect to the original one.
From basic thermodynamics we know that
this entropy flux is $\beta\,{\cal Q}$, where ${\cal Q}$ is
the heat flowing to the thermal environment, that is:
minus the change in potential energy of the system minus the work
performed on the system.  The
reason for \eqref{ef} is the physical condition of local detailed
balance, which ensures that the
ratio of probabilities of the forward with respect to the backward
trajectory is given by the exponential of the entropy flux in the forward
trajectory~\cite{entr}.

The time-symmetric part  $T(\traj)$ of the action $A$
is the excess in activity, having an essential
role in dynamical fluctuation theory for the large deviations of the occupations,
and in nonequilibrium studies  of phase transitions~\cite{mnw,DA1}.
Being in linear response theory (small $h_s)$, we are actually more concerned with
the first order in excess activity $\tau(\traj,s) = \frac{\delta}{\delta h_s} T(\traj)|_ {h_s=0}$.
The excess in activity quantifies how ``frenetic'' is the motion in the perturbed process,
compared with the unperturbed one.
For example, for a Markov jump process with transition rates
$W(x\to y)$ between states $x\rightarrow y$, the traffic
is equal to twice the time-integrated escape rates over a trajectory,
$2 \int_0^t\,\id s\, \sum_y W(\omega_s\to y)$
and its excess  $\tau(s)$
(here $\tau(\traj,s) = \tau(s)$ depends only on the state $\omega_s$)
to linear order in $V$ is
\begin{eqnarray}\label{3}
\tau(s)=2\sum_y W(\omega_s\to y)\left\{e^{\frac{\beta}{2}[V(y)-V(\omega_s)]}-1\right\}&&\\
\simeq
 \beta \sum_y W(\omega_s\to y)[V(y)-V(\omega_s)]
&=& \beta \frac{\id V}{\id s}\nonumber
\end{eqnarray}
The last expression is of course formal, the mathematical meaning being 
$\id V/\id s = LV(s)$.

Since $S(\traj)$ is already linear in $h_s$, we have
\begin{equation}\label{exp}
 -A(\traj) = \frac{1}{2}S(\traj)
-\frac 1{2} \int_0^t \id s\,h_s\,\tau(\traj,s) + O(h_s^2)
\end{equation}
We are now ready to derive the FDR for a single-time observable $O(\traj)
= Q(t)$ at time $t>0$.
Its expectation in the linear response regime is
$\mean{Q(t)}^h = \mean{Q(t)} +\int_0^t\id s\, h_s R_{QV}(t,s)$, which
we can rewrite with (\ref{Ph}) as
$\mean{Q(t) e^{-A(\traj)}} - \mean{Q(t)} \simeq - \mean{Q(t) A(\traj)}
= \int_0^t\id s h_s R_{QV}(t,s) $.
With \eqref{ef} and \eqref{exp}, it is thus straightforward to see
that the  response function is equal to
\begin{equation}\label{mainid}
 R_{QV}(t,s) = \frac{\beta}{2}\frac{\id}{\id s}\left< V(s)Q(t)
 \right>- \frac 1{2}\left<\tau(\traj,s)Q(t)\right>
\end{equation}
which, as we will see, is a nonequilibrium FDR that
generalizes and unifies previous formulations.
We stress that the derivation of \eqref{mainid}
is independent of many details of the dynamics and hence of its
generator $L$: we went
beyond formalities of a perturbative first-order calculation by
specifying the origin of the two parts on the right-hand side of
\eqref{mainid} in terms of (i) the entropy flux and (ii) the dynamical
activity~\cite{mnw,DA1,t-symm}, in excess by the time-dependent perturbation.

When $\tau(\omega,s) = \tau(s)$, as in \eqref{3} or for overdamped diffusions \cite{note2}, formula
(\ref{mainid}) simplifies to
\begin{equation}\label{nonefd}
R_{QV}(t,s) = \frac{\beta}{2}\,\frac{\id}{\id s}\left<
V(s)\,Q(t)\right> - \frac{\beta}{2}\,\left<\frac{\id}{\id s} V(s)
\,Q(t)\right>
\end{equation}
The right-hand side of \eqref{nonefd} is nonzero because of the
correlation of $V(s)$ with $Q(t)$ for $t>s$~\cite{note_noneq}.

When the unperturbed dynamics is 
detailed balanced, we can recover (\ref{efd}) as a special
case of (\ref{nonefd}) by using 
time-reversal symmetry:
for
$s<t$, $\langle \id V/\id s \,Q(t)\rangle_{\textrm{eq}} =$ $
\langle \id V/\id t \,Q(s)\rangle_{\textrm{eq}} = $  $\id/\id
t\,\langle V(t) \,Q(s)\rangle_{\textrm{eq}}$, where the last equality
is a consequence of the Markov property of the dynamics. Because of stationarity
this is then also equal to $- \id/\id
s\,\langle V(s) \,Q(t)\rangle_{\textrm{eq}}$, so that in equilibrium
the second term of (\ref{nonefd})  gives exactly the same contribution 
as its first term, and (\ref{efd}) is recovered.


As an illustration of the general formula,  consider
 a system with a flow of interacting particles due to the presence of
reservoirs at different chemical potentials.
We choose both $Q=V={\cal N}$, where ${\cal N}$ is the particle number.
Hence,  $h_s$ induces a global shift in chemical potentials,
and the linearized excess activity $\tau(s)=\beta \id{\cal N}/\id s = 
\beta {\cal J}(s)$
is proportional to the systematic particle current ${\cal J}(s)$
into the system via its boundaries, i.e.~to the expected 
current given the configuration at time $s$.
Numerically it is convenient to test
the time-integrated version of \eqref{nonefd} for constant
perturbation $h_s=h$ for $s\ge 0$,
\begin{equation}\label{chi}
\chi(t) / \beta = [ C(t) + D(t) ]/2
\end{equation}
with susceptibility
$\chi(t) = (\mean{{\cal N}(t)}^h - \mean{ {\cal N}})/h$,
correlation function
$C(t) = \langle {\cal N}^2\rangle -  \langle {\cal N}(0){\cal N}(t)\rangle$,
and nonequilibrium  term
$D(t) = - \int_0^t\id s \mean{ {\cal J}(s){\cal N}(t) }$.
Again, \eqref{chi} holds irrespective of the details of interaction or driving.

\begin{figure}[!bt]
\begin{center}
\includegraphics[angle=0,width=8.2cm]{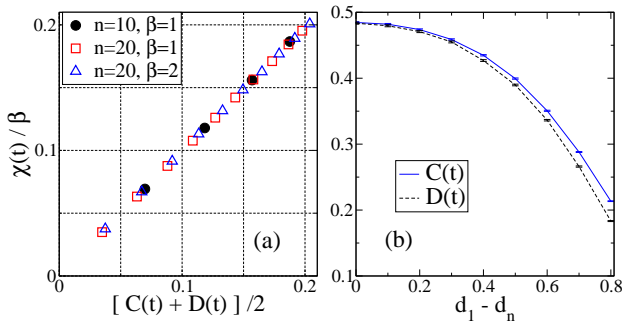}
\end{center}
\caption{(Color online) (a) Plots of $\chi(t)/\beta$ vs $[C(t)+D(t)]/2$ for
times $t=5$, $10, \ldots, 50$ (other parameters: see legend, and
$h=-0.1$, $d_1-d_n=0.8$).
(b) $C(t=50)$ and $D(t=50)$ as a function of the unbalance 
between reservoir densities, for $\beta=1$, $n=20$. 
At equilibrium ($d_1=d_n$) one has  $C(50) = D(50)$ as expected, while
the difference between $C(50)$ and $D(50)$ is not negligible
for strong nonequilibrium.
}
\label{fig}
\end{figure}

Here we simulate a boundary driven one-dimensional exclusion process,
a simple paradigmatic model for transport.
A state $x$ is an array with empty sites $x^i=0$ and particles
$x^i=1$.  Nonequilibrium is imposed
at boundary sites $i=1$ and $i=n$, which are in contact with
reservoirs with particle densities $d_1$ and $d_n=1-d_1$, respectively.
There is a nearest neighbor attraction with energy
$H(x)  = -\sum_{i=1}^{n-1} x^i x^{i+1}$ 
determining the Kawasaki dynamics in the bulk.
In this context the systematic current arises from two
possible transitions: $x\to y$, with $y^1=1-x^1$, and
$x\to z$, with $z^n=1-x^n$.
Hence \[
{\cal J}(x)\! =\!
[{\cal N}(y) - {\cal N}(x)] W(x\!\shortrightarrow\!y) +
[{\cal N}(z) - {\cal N}(x)] W(x\!\shortrightarrow\!z)
\]
where e.g. $W(x\to y)\propto d_1 e^{-\beta[H(y)-H(x)]/2}$
if a particle enters at $i=1$.
Examples in Fig.~\ref{fig}(a) visualize Eq.~(\ref{chi}),
while the magnitude of $C(t=50)$ and $D(t=50)$ is shown in Fig.~\ref{fig}(b)
as a function of the driving $d_1-d_n$.
Far from equilibrium $C(t) \ne D(t)$ and thus the correlation 
function $C(t)$ alone
cannot provide a good estimate of the susceptibility $\chi(t)$. 

In order to connect Eq.~\eqref{mainid} [or \eqref{nonefd}]
with some previous formulations, it is easy enough to rewrite it in 
another form. First remember that
causality implies $R_{QV}(s,t)=0$ for
$s\le t$.  Hence, for \eqref{mainid} to be valid for all $s,t$ one needs
\begin{equation}\label{cau}
\beta\frac{\id}{\id t}\langle V(t)Q(s)\rangle =
\langle\tau(\omega,t)Q(s)\rangle\quad\textrm{for}\;  s \le t\;,
\end{equation}
which is automatically satisfied in our framework: it 
follows from $\big<e^{-A(\traj)}\big>=1$, equivalent to
$\mean{A} = 0$ to first order in $h_s$, via partial integration of
\eqref{ef}. By subtracting \eqref{cau} from \eqref{mainid} one finds
 \begin{equation}\label{responsesym}
\begin{split}
  R_{QV}(t,s) =& \,
  \frac{\beta}{2} \left( \frac{\id}{\id s}\mean{ V(s)Q(t)}
                       - \frac{\id}{\id t}\mean{ V(t)Q(s)} \right)\\
&- \frac 1{2}\mean{ \tau(\omega,s)Q(t)-  \tau(\omega,t)Q(s) }
\end{split}
 \end{equation}
In stationarity the second line in \eqref{responsesym} is an extra term compared 
to the equilibrium version \eqref{efd}, in which the time-antisymmetric correlation
between activity $\tau$ and the observable $Q$ is evaluated.
This term has been called {\em asymmetry} in
studies of overdamped
Langevin equations~\cite{ckp} and of discrete stochastic systems~\cite{lipp,diez}.

Again, when $\tau(\omega,s) = \tau(s)$, as in (\ref{3}) and (\ref{nonefd}), equation 
(\ref{cau}) implies that $\tau(s) = \beta\,LV(s)$.
Hence \eqref{mainid} turns into
\begin{equation}\label{marmainid}
 R_{QV}(t,s) =
 \frac{\beta}{2}\frac{\id}{\id s}\left< V(s)Q(t) \right>-
 \frac {\beta}{2}\left<LV(s)\,Q(t)\right>
\end{equation}
which is the rigorous version of~\eqref{nonefd}.
Let us now make contact with some previous formulations for 
stationary processes, where   
$\frac{\id}{\id s} \left<V(s)\,Q(t)\right> = 
\frac{\id}{\id s}  \left<V(0)\,Q(t-s)\right> = 
-\frac{\id}{\id t} \left<V(s)\,Q(t)\right> $.
From our definitions, the last term equals $-\mean{V(s) LQ(t)}$.
Alternatively we can think of $Q(t)$ correlated with $V(s)$ evolved backward
in time by the adjoint  $L^*$ (which generates the time-reversed process): 
$\mean{V(s) LQ(t)} = \mean{L^* V(s) Q(t)}$.
Eq.~\eqref{marmainid} can then be rewritten as
\begin{equation}\label{revj}
  R_{QV}(t,s) = -\frac{\beta}{2}
\langle L^* V(s)\,Q(t) + L\, V(s)\,Q(t)\rangle
\end{equation}
which is a generalization of Eq.~(2.15) in \cite{jona}.  Indeed,
in the context of fluctuation theory around diffusive scaling limits,
the $L^*$ is referred to as the adjoint hydrodynamics
in the infinite-dimensional treatment in Section 2.3 of \cite{jona}.
It is emphasized there that the response for the
adjoint process is typically along the reversed trajectory from a
spontaneous fluctuation in the original dynamics.
This is also explicit in \eqref{revj},
as the response in the time-reversed process amounts to the same expression upon
exchanging $L$ with $L^*$ and $t$ with $s$.
Note however that one needs the stationary distribution to know $L^*$. 
The knowledge of this distribution enables also the derivation of other
FDRs~\cite{mprv,deker,vulp,hanggi,ss2009}.
On the other hand, Eq.~\eqref{marmainid}
does not involve the stationary law, except for the statistical averaging.

We also recover the interpretation of \cite{gaw}:
by using \eqref{revj}, we rewrite the stationary version of \eqref{mainid} as
\begin{equation}\label{markovresponse2}
  R_{QV}(t,s) \!=\!
   \,\beta\frac{\id}{\id s}\!\left< V(s)Q(t)\right>
 - \frac{\beta}{2}\left<[\big(L\!-\!\!L^*\big)V](s)\,Q(t)\right>
 \end{equation}
Note that the response regains its equilibrium form
[only the first term in the right-hand side of \eqref{markovresponse2}]
whenever $V$ is a time-direction neutral observable, in the precise
 sense that $LV = L^*V$.  
Furthermore one can check
(say, for overdamped diffusions and for jump processes) that
$L- L^* = 2\,\frac{j}{\rho}\cdot\nabla$, where $j$ is the
probability current and $\rho$ 
is the stationary distribution on the states. 
Their ratio is a drift velocity $v = j/\rho$.
Hence, for a time-direction neutral potential $V$, the probability
current is orthogonal to its gradient, $v\cdot \nabla V=0$, and
the equilibrium form of the response is obtained, in
agreement with the observations in \cite{sasa}.
The same effect is achieved by describing the system in the Lagrangian
frame moving with drift velocity $v$.  The second term
in \eqref{markovresponse2} vanishes also in this case, and 
\eqref{markovresponse2} yields exactly the interpretation 
(and Eq.~4.5) of~\cite{gaw},
which was recently experimentally verified~\cite{cili}. 

To finish, we revisit the relation with dissipation to confirm the
prediction in~\cite{house} that the usual equilibrium
relation between response and dissipation is preserved when taking
into account only the {\it excess} heating and ignoring the
housekeeping heat. 
During a specific trajectory of the system, the heat dissipation
can be split into two parts: ${\cal Q}_{hk}+{\cal Q}_{ex}$~\cite{house}.
The first term is the housekeeping heat, which is the 
heat produced in the unperturbed dynamics. This is the heat that
drives (or is the result of) the system out of equilibrium. 
The second part: ${\cal Q}_{ex}$, is the extra heat generated through the perturbation,
and is just $1/\beta$ time the excess entropy production,
which we already encountered in our calculations:
\begin{eqnarray}
 {\cal Q}_{ex} \!\!&=& \frac{1}{\beta}\left<S(\traj)\right>^h
=\int_0^t\id s\,h_s\,\frac{\id}{\id s}\left< V(s) \right>^h
\end{eqnarray}
This expression is of the same form as in equilibrium, 
so we will not repeat the calculations here, only the conclusion:
the (excess) heat is given by the imaginary
part of the Fourier transform of $R_{VV}$, or equivalently,
the Fourier transform of the time-antisymmetric part of $R_{VV}$.
Moreover, we have already written down this time-antisymmetric part: 
it is the right-hand side of \eqref{responsesym}, extended  to $t\le s$.

In conclusion, from physical constraints on the probability of
trajectories, we have obtained a general FDR for the response of a
driven system to the addition of a potential.  It lifts the nonequilibrium FDR
beyond formal first order perturbation theory applied to a specific dynamics,
by identifying in general physical terms the statistical quantities that
determine the response.
Previous formulations are recovered as specific cases.
Finally, we have given observational significance to
the notion of dynamical activity,
which so far has mostly appeared as a theoretical concept in
fluctuation theories.

\paragraph*{Acknowledgments:}
We are grateful to  W.~De Roeck, K.~Gawedzki, and K.~Neto\v{c}n\'{y}
for enlightening discussions.
B.W.~is an aspirant of FWO, Flanders.
M.B.~benefits from K.U.Leuven grant OT/07/034A.

\bibliographystyle{plain}

\begin{thebibliography}{10}
\bibitem{note2} We do not discuss here some extra difficulties 
for underdamped diffusions, but it is valid for Markov jump
processes and for overdamped diffusions.
\end{thebibliography}

\end{document}